\documentstyle[epsf,psfig,preprint,aps]{revtex}
\tightenlines

\def\EQ{\begin{equation}}
\def\EN{\end{equation}}
\def\EQA{\begin{eqnarray}}
\def\ENA{\end{eqnarray}}

\begin{document}
\title{  
Logarithmic corrections to scaling in turbulent thermal convection}
\author{ B. Dubrulle$^{1,2}$}

\address{
$^1$ CNRS, Service d'Astrophysique, CE Saclay, F-91191 Gif sur Yvette Cedex, France\\ 
$^2$ CNRS, URA 285, Observatoire
Midi-Pyr\'en\'ees, 14 avenue Belin, F-
31400 Toulouse, France
}
\maketitle

\begin{abstract}
 We use an analytic toy model of turbulent convection to show that
most of the scaling regimes are spoiled by logarithmic corrections, in a way consistent with the most accurate experimental measurements available nowadays. This sets a need for the search of new measurable quantities which are less prone to dimensional theories.
\end{abstract}

\begin{flushleft}
47.27 -i Turbulent flows, convection and heat transfer\\
47.27.Eq Turbulence simulation and modeling\\
47.27.Te Convection and heat transfer 
\end{flushleft}

\vspace{0.1cm}

\narrowtext
When an horizontal layer of fluid is heated from below, a heat exchange from the bottom to the top layer occurs. Dimensional consideration show that the non-dimensional heat exchange $Nu$ depends on the geometry (via for example the aspect ratio), on the boundary conditions, and on the Rayleigh $Ra$ and the Prandtl number $Pr$. Further dimensional analysis of the dynamical equations governing the convective layer suggests that this dependence be a power-law. The exponent of the power law depends on the (dimensional) basic hypothesis. The classical theory predicts that $Nu\sim Ra^{\beta}$ with $\beta=0.33$, which seems to be observed in electro-chemical convection  \cite{Gols}. Other theories lead to $\beta=0.2$ or $0.24$ \cite{CCS97}, $\beta=2/7$  \cite{Cast89,ShraSigg90} or $\beta=1/2$ \cite{Krai62} depending on the regime considered. A unifying synthesis of the possible scaling regimes, including their Prandtl dependence has been recently made by Grossman and Lohse \cite{GrosLohs00,GrosLohs01}. From the experimental point of view, the situation is rather unclear, since almost all the range of exponents between 0.25 and 0.33 has been measured (see \cite{GrosLohs00} for a recent detailed review of the experimental measurements). Furthermore, recent experiments with fluids subject large variation of the Prandtl number have led to a new exploration of the phase parameter, and uncovered new scaling laws as a function of the Prandlt number. This new exploration can be used as a stringent tool to discriminate between the various theories, since two competing theories can provide the same value of $\beta$, but different Prandtl number dependence.\

On the other hand, it is not quite clear whether the present dimensional theories satisfactorily capture the dependence of $Nu$ as a function of $Pr$ and $Ra$. In a recent serie of experiments, conducted with acetone at various aspect ratio, \cite{XBA00} show that their measurements are inconsistent with a single power law $Nu(Ra)$, because the local effective exponent $\beta_{eff}=d\ln Nu/d\ln Ra$ varies continuously with Ra in the range of Rayleigh number they consider. This effect can easily be accounted for in the scaling theory by considering a superposition of scaling laws, as proposed by \cite{GrosLohs00}. There is another solution, connected with the existence of logarithmic corrections to scalings (see e.g. \cite{ShraSigg90}). The difficulty associated with such a solution is that logarithmic corrections cannot be uncovered by dimensional analysis (nor numerical simulations!), and require detailed analytical computations. At the present time, they are no available analytical solutions of the full Boussinesq equations describing the dynamics of the convective motions, in the turbulent regime. Some time ago, Malkus \cite{Malk54} proposed to combine a weakly non-linear theory past the critical point, combined with a maximum principle to obtain approximate analytical solutions of the full problem. This approach developed further by Howard, Roberts, Stewartson and Herring (see refs \cite{Buss78} for a summary) gives $Nu=0.24 (Ra \ln Ra)^{3/10}$. Such a law with logarithmic correction fits the recent experimental data by Niemela et al. \cite{NSSD00} very well.\

In the present work, we explore the predictions obtained using a solvable model of turbulent convection.
This model couples large scale
mean sheared velocity and temperature fields $U=(U(z),0,0)$ and 
$\Theta(z)$, with small scale random velocity and temperature fields.
This kind of large scale geometry is generally accepted as representative
of the Boussinesq convection in the boundary layers, due to the shearing effect 
of the large convective cells.
The model is closed by deriving an equation for the 
random component from the Boussinesq equation using two simplifying 
assumptions: i) the non-linear
interactions of the small scale scales 
between themselves is modeled via a turbulent viscosity; ii) the small
scale generation via the breaking of large scale structures 
is modeled by a random small scale forcing with prescribed statistics.
This results in a linear 
stochastic equation for the random small scales, which can be 
analytically solved in the shear flow geometry by a decomposition of
the small waves into localized wave-packets. In \cite{DLS00,Dubr00}, the solution of the turbulent model were computed in the restricted case of 2D geometry, and turbulent Prandtl number equal to unity. This leads to expression of the mean and fluctuating velocity and temperature profile, as a function of $Nu$, $Ra$ and $Pr$. In \cite{Dubr00}, these results were combined with dimensional estimates to derive the scaling laws governing the transport of heat. In the present contribution, we proceed one step further and use the analytical predictions about the profiles to compute directly the heat and energy dissipated across the turbulent boundary layer. From this, we derive an analytical expression of the non-dimensional heat flux, as a function of $Pr$ and $Ra$.\

There can be a debate about whether such a simple model correctly accounts for the complex dynamics of the full, turbulent Boussinesq equations. A systematic numerical check  of the various hypotheses pertaining the model has been undertaken in the past years \cite{DLSW00} and is still underway. In any case, we think that this model is very illustrative as to what kind of mechanism could take place in real, turbulent convection. Our computations show that most of the likely scaling regimes are spoiled by logarithmic corrections, in a way consistent with the most accurate experimental measurements available nowadays. We believe that this sets a need for the search of new measurable quantities which are less prone to dimensional theories.\

To set up notations, we start by a brief summary of the turbulent model \cite{DLS00,Dubr00} and of results obtained within it. In this model, the dynamics of the turbulent boundary layer is obtained from solutions of two coupled sets of equations. The first one described the dynamics of the mean (shear like) velocity ${\bf U}=<{\bf u}>=(U(z),0,0)$ and temperature $\Theta(z)=<\theta>$:
\EQA
\partial_t U+\partial_z<u'w'>
&=&-\partial_x P 
+Pr\partial_{z}^2 U,\nonumber\\
\partial_t \Theta
+\partial_z<w'\theta'> 
&=&
\partial_z^2\Theta.  
\label{LSEq}
\ENA
These equations have been  expressed, following \cite{ShraSigg90} in units the thermal diffusivity
and the cell height.
Here, $Pr$ is the Prandtl number, the primes denote fluctuating (small scale) quantities and $<>$ the
averaging. In the stationary case, we get from (\ref{LSEq}) that $\partial_x P$
is a constant, independent of $z$. In the laminar case where 
$<u'w'>$ and $<w'\theta'>$ are negligible, we thus obtain the well known
parabolic profile for the velocity and the linear profile for the temperature.
In the turbulent case, the profiles are linear within the thermal or the
viscous layer, while outside this layer, they are given by the condition 
\EQ
\partial_z <u' w'>=-\partial_x P, \quad \partial_z <w'\theta'>=0.
\label{turbulentcase}
\EN   
To close the system, we need $<u'w'>$ and $<w'\theta '>$. They are obtained as solution of a linear, stochastic equation valid for localized wave-packets of velocity and temperature:
\EQA
D_t \hat u_i
&=&-ik_i \hat p- \hat w \partial_z U \delta_{i1}
+Ra Pr \hat \theta\delta_{i3}
-Pr^t {\bf k}^2 \hat u_i+\hat f^{(u)}_i\nonumber\\
D_t \hat \theta
&=&-\hat w\partial_z \Theta 
-{\bf k}^2\hat\theta+\hat f^{(\theta)},
\label{ssEqmod}
\ENA
where
\EQ
{\hat u}({\bf x},{\bf k},t)=\int g(\vert
{\bf x-x'}\vert)e^{i{\bf k\cdot (x-x')}}
{\bf u}({\bf x'},t)d{\bf x'},
\label{gabordef}
\EN
$g$ being a function which decreases rapidly at infinity.
We have dropped primes on fluctuating quantities for convenient notations
and introduced the total derivative $D_t=
\partial_t +U\partial_x
-\partial_z (U k_x) \partial_{k_z}$ .
Note that the linear part of (\ref{ssEqmod}) is exact and describes
 the non-local interactions
between the mean and the fluctuating part. The major approximation of the model is to lump the non-linear terms describing local interactions between fluctuations
into a  turbulent 
viscosity, or equivalently, into a turbulent 
Prandtl number $Pr^t$. The forces $f^{(u)}$ and $f^{(\theta)}$ appearing in (\ref{ssEqmod}) are small scale random forces which are introduced to model the seeding of small scales by energy cascades (for example via turbulent plumes, detaching from the wall).\

The analytical solutions of (\ref{LSEq}) and (\ref{ssEqmod}) have been obtained in \cite{DLS00,Dubr00} in the 2D case (no movement in the direction transverse to the mean flow), and for $Pr^t=1$. There are numerical  and analytical evidence that 
2D geometry is sufficient to capture the physical mechanism responsible 
for the $Nu$ scaling with $Ra$ \cite{Wern93} and also to capture
the correct shape of the equilibrium profile in the neutral case\cite{Naza99}. Note that
the vortex stretching, which is theoretically absent in 2D 
geometry, has been implicitly accounted for via the turbulent
viscosity. The unit value of $Pr^t$ was dictated by our ignorance of the exact value of this parameter (which is likely to depend on $Pr$). We shall keep in mind that it may induce a wrong dependence of the correlations with respect to the Prandtl number. In any case, we shall consider these solutions as a toy model of turbulent thermal convection, and investigate their scaling properties. For the mean flows, the solution are \cite{Dubr00}:
\EQ
\partial_z U\sim \frac{1}{z},\quad \partial_z \Theta \sim \frac{1}{z^2},
\label{solutionprof}
\EN
resulting in a constant (with $z$) Richardson number: 
\EQ
R_i=Ra Pr\frac{\partial_z \Theta}{\left(\partial_z U\right)^2}.
\label{ridchardson}
\EN
In developed convective turbulence, this number is
large and negative $R_i\ll -1$. This motivates a large $Ri$ expansion of the solution of (\ref{ssEqmod}), which provides the expression of the correlations as :
\EQA
<w' u'>&\approx& \frac{u_\ast^2}{\partial_z U}\propto z,\nonumber\\
<w'\theta'>&\approx& \frac{u_\ast^2 \sqrt{-R_i}}{RaPr}\propto cte=Nu,\nonumber\\
<w'^2>&\approx& \frac{u_\ast^2}{\partial_z U},\nonumber\\
<\theta'^2>&\approx& R_i \frac{u_\ast^2\partial_z U}{(RaPr)^2}=\frac{u_\ast^2\partial_z \Theta}{RaPr\partial_z U}.
\label{oufi}
\ENA
where we have introduced $u_\ast^2=(\partial_z U/k_\ast^2)^{2\sqrt{-R_i}/3}$. Note that the last equation of (\ref{oufi}) just reflects the balance between the vertical energy and the buoyancy force. Such a balance can be expected to hold only at $Pr>1$ \cite{CCS97}. For low $Pr$, temperature fluctuations were found to behave like a passive scalar, except in the boundary layers, or in the center of the cell \cite{SNS98}. Since our expression pertain only to the boundary layers, we feel safe to consider it for both the $Pr>1$ and the $Pr<1$ regime.\

The second equation of (\ref{oufi}) can be used to express $u_\ast$ as a function of $Nu$ and $Ri$. Then, we can eliminate $Ri$ by
matching the 
turbulent profiles with the viscous or diffusive solutions
$Pr\partial_z U=u_\tau^2$ and $\partial_z \Theta=Nu$, where $u_\tau$ is the 
friction velocity and $Nu$ the Nusselt number. For this, we introduce the thermal length scale $\lambda_\theta=1/2Nu$ and 
the viscous length scale $\lambda_u=K Pr/u_\tau$ ($K\approx 0.4$ is the Karman constant), and operate the matching via smooth functions according to:
\EQA
\partial_z U=\frac{u_\tau^2 Pr^{-1}}{\sqrt{1+(z/\lambda_u)^2}},\nonumber\\
\partial_z \Theta =\frac{Nu}{1+(z/\lambda_\theta)^2}.
\label{solutionsmooth}
\ENA
The shape of $\partial_z U$ is dictated by the exact analytical solutions of the profile found by \cite{Naza99} in a neutral layer. The shape of $\partial_z \Theta$ was chosen as the simplest smooth function matching the viscous and turbulent layers. Using these expressions, we can eliminate $Ri$ and write for the fluctuations:
\EQA
<w'^2>&=& Nu^{3/2}(Ra Pr)^{1/2}\frac{u_\tau}{\partial_z U},\nonumber\\
&=& Nu^{3/2}(Ra Pr)^{1/2}\frac{Pr}{u_\tau}\sqrt{1+(z/\lambda_u)^2},\,\nonumber\\
<\theta'^2>&=& u_\tau\frac{Nu^{3/2}}{(RaPr)^{1/2}}\frac{\partial_z \Theta}{\partial_z U}\nonumber\\
&=& \frac{Pr}{u_\tau}\frac{Nu^{5/2}}{(RaPr)^{1/2}}\frac{\sqrt{1+(z/\lambda_u)^2}}{1+(z/\lambda_\theta)^2}.
\label{oufo}
\ENA
Note that these fluctuations exist only for $z>\lambda_u$. The expression we derived are only valid inside the turbulent layers, which we define as the locations were the fluctuations of horizontal velocity and of temperature reach a maximum. The scales at which these maxima are reached are labeled $\lambda_{BLV}$ and $\lambda_{BLT}$, respectively. Beyond these scales, the flow enters the "bulk region", in which the velocity and temperature fields are organized into convective large scale, whose details are non-universal and strongly depend on the geometry (aspect ratio) of the cell. It is difficult to evaluate analytically the vertical dependence of the profiles. On the other hand, support from  data can only be provided by numerical simulations, encompassing relatively low Rayleigh numbers. One important outcome of these simulations \cite{VerzCamu99,Kerr96,KerrHerr00} is the existence of two distinct flow regimes: one for $Pr<0.35$ in which the flow is dominated by a large re-circulation cell, and one $Pr>0.35$ in which isolated thermal plumes can develop. The detailed scaling of the velocity pattern has only been partially studied by Kerr \cite{KerrPC} in the regime $Pr=0.7$,  $Ra=8\times 10^7$ and for aspect ratio 1. He found no indication of the logarithmic region, but rather a variation of the horizontal velocity profile consistent with a $z^{-1/2}$ law. This can be easily understood by noting first, that in this Prandtl regime, the Reynolds number corresponding to $Ra=8\times 10^7$ is too low to allow for a well developed turbulent boundary layer. Therefore, the logarithmic region is non-detectable. On the other hand, in the core, there is a tendency for strongly homogeneous temperature (due perhaps to the presence of plumes, which favor the good mixing), with nearly constant value $\Theta_0$ throughout the central region. The classical free fall velocity argument then predicts vertical velocities varying like $W^2\sim Ra Pr \Theta_0 z$, resulting, by incompressibility $\partial_z U\sim \partial_z W$ in a $z^{-1/2}$ law for the horizontal profile. Given the small extent of the logarithmic zone in this regime, we shall therefore approximate the bulk velocity in this region, at $Pr>0.35$ by $u_\tau (z/\lambda_u)^{-1/2}$ (to ensure the matching with the viscous layer).
Unfortunately, no comparable measurements are available at low Pr. On the one hand, we can expect the logarithmic region to be more extended in this regime, because of the larger Reynolds number involved. Interestingly enough, inversion of the temperature profiles in the central region have been detected in the experiment of \cite{SNS98}, with a nearly constant and positive value of $dT/dz$ (whether this is correlated with the absence of plumes is an interesting open question). In such a case, the free-fall argument (which is valid provided we are in the Bolgiano regime, i.e. at large enough scales) predicts vertical velocities varying like $W^2\sim Ra Pr z^2$, i.e., via the incompressibility, constant horizontal velocity profiles. Flatter profiles of horizontal velocity have indeed been observed in some numerical simulations of Kerr \cite{KerrPC}, but the scaling has not been properly checked. In any case, as a first approximation, we shall assume that the horizontal velocity in the central region and in the $Pr<0.35$ regime, is constant $U_0$. The matching of this constant with the logarithmic profile then imposes $U_0=Ku_\tau(\ln(\lambda_o)+B)$, where $\lambda_o$ is the outer scale, and $B$ is a constant which may depend on the Prandtk number (under neutral condition, $B\approx 5$). In the sequel, we assume $U_0=f u_\tau$, and assume that $f$ is independent of the Prandtl number (for $Pr<0.35$).\

These estimates can be further used to determine the scaling behaviors of the two scales $\lambda_{BLT}$ and $\lambda_{BLV}$ characterizing the turbulent boundary layer. For this, we follow the logic of our turbulent model which states that in the boundary layer, the fluctuations are 
passively advected by the large scale
horizontal velocity and obeys $U\partial_x u'= Pr \partial_z^2 u'$ or  $U\partial_x \theta'=\partial_z^2 \theta'$. On dimensional ground, this shows that the two scales are determined via the two implicit equations:
\EQ
{U(\lambda_{BLT})}\sim \frac{1}{\lambda_{BLT}^2},\quad {U(\lambda_{BLV})}\sim \frac{Pr}{\lambda_{BLV}^2}.
\label{lengthscales}
\EN
Here, we have ignored the aspect ratio dependence. Aspect ratio obviously affects the scaling regimes, but since we cannot quantify its effect in the large scale circulation, we shall ignore it hereafter.
Depending on the Reynolds number $Re=u_\tau/Pr$ and on the Prandtl number, different regimes of velocity will be probed by this implicit relations, and the scaling of $\lambda_{BLV}$ and $\lambda_{BLT}$ as a function of $Pr$ and $u_\tau$ will vary. Introducing $\epsilon(\lambda)=d\ln U/d\ln z\vert_{z=\lambda}$  (where $\lambda=\lambda_{BLV}$ or $\lambda_{BLT}$), we can summarize this dependence as:
\EQA
\lambda_{BLT}&=&Pr^{\epsilon/(2+\epsilon)}u_\tau^{-(1+\epsilon)/(2+\epsilon)}\left(\frac{1}{\ln\lambda_{BLT}}+\frac{1}{f}\right)^{\delta(\epsilon)/(2+\epsilon)},\nonumber\\
\lambda_{BLV}&=&\left(\frac{Pr}{ u_\tau}\right)^{(1+\epsilon)/(2+\epsilon)}\left(\frac{1}{\ln\lambda_{BLV}}+\frac{1}{f}\right)^{\delta(\epsilon)/(2+\epsilon)}.
\label{resume}
\ENA
Here, $\delta(\epsilon)$ is equal to 1 if $\epsilon=0$ and zero elsewhere.
The dependence of $\epsilon$ in $\lambda_{BLT}$ and $\lambda_{BLV}$ has been omitted for simpler notations. Also, the logarithmic and constant velocity regime have been lumped into the single notation $U(\lambda)\sim u_\tau/(1/\ln\lambda+1/f)$ which patches the two consecutive characteristic behaviors.\

These considerations can be used to compute the Rayleigh and Prandtl dependence of the Nusselt number. For this, we follow \cite{ShraSigg90,GrosLohs00} and use the two rigorous relations involving the global average of the kinetic and thermal dissipation rates:
\EQA
<Pr(\partial_i u_j)^2>_V&=&Ra Pr (Nu-1),\nonumber\\
<(\partial_i \theta)^2>_V&=&Nu.
\label{exacts}
\ENA
In (\ref{exacts}) averages are taken over spatial volumes. Dissipation takes place both in the boundary layers, and in the "bulk" region. However, several independent observations militates in favor of the importance of boundary layers, rather than bulk region, to determine the scaling of the Nusselt with the Rayleigh number. For example, Julien et al \cite{WernPC} observe that when switching from no-slip to stress-free boundary conditions, the exponent of the power-law Nusselt versus Rayleigh changes from close to $2/7$ to close to $1/3$. More recently, Ciliberto and Laroche \cite{CiliLaro99} observed that both the prefactor and the exponent of the power-law changes when switching from smooth to rough bottom plate in the convective cell. On the other hand, blocking the large scale circulation does not affect the heat transport \cite{CCL96}. In numerical simulations at $Pr=0.7$ and $Ra=8\times 10^7$, Kerr \cite{KerrPC} noted that a change of aspect ratio (which changes the large scale circulation) only affects the energy dissipation in the bulk of the flow. In the boundary layers, the total dissipation seems to remain always about 1/4 of the total dissipation. Gathering all these information, we shall assume that a constant fraction of heat  and energy  is dissipated in the boundary layers, independently of geometry, Rayleigh number, Prandtl number... This allow us to consider similar equations to (\ref{exacts}), where the average is only taken over the volume encompassed by the boundary layers. The advantage is that, in that region, we can use our exact analytical solutions to compute the kinetic and thermal dissipation.\

In each case, these dissipation include two contributions: one from the mean flow, and one from the fluctuations. For the mean flow contribution, we use (\ref{solutionsmooth}) and find:
\EQA
\epsilon_U&=&Pr\int_{0}^{\lambda_{BLV}} dz(\partial_z U)^2=K u_\tau^3 arctg\left(\frac{\lambda_{BLT}}{\lambda_u}\right),\nonumber\\
\epsilon_\Theta&=&\int_{0}^{\lambda_{BLT}} dz(\partial_z \Theta)^2=
\frac{Nu^2}{2\lambda_\theta}\left[\frac{\lambda_{BLT}/\lambda_\theta}{1+(\lambda_{BLT}/\lambda_\theta)^2}+arctg\left(\frac{\lambda_{BLT}}{\lambda_\theta}\right)\right].
\label{meandissi}
\ENA
For the fluctuations, we first note that the velocity and thermal fluctuations are non-negligible only for $z>\lambda_u$ \footnote{Temperature fluctuations are generated through velocity fluctuations, and, thus, are slaved to them}. Further, we use incompressibility to get $(\partial_z u)^2\sim \lambda_{BLV}^{-2}(\partial_x u)^2\sim \lambda_{BLV}^{-2}(\partial_z w)^2$. Then, we approximate $\partial_z w\sim \partial_z \sqrt{<w^2>}$ and $\partial_z \theta\sim \partial_z \sqrt{<\theta^2>}$. Using (\ref{oufo}) and performing the integration, we finally obtain for the contribution of the fluctuations:
\EQA
\epsilon_u&\sim &\frac{(NuPr)^{3/2} Ra^{1/2}}{4K}\left(\frac{1}{2}-\frac{(\lambda_{BLV}/\lambda_u)}{\sqrt{1+(\lambda_{BLV}/\lambda_u)^2}}+\ln\left(\frac{2\sqrt{1+(\lambda_{BLV}/\lambda_u)^2}+2(\lambda_{BLV}/\lambda_u)}{2\sqrt{2}+2}\right)\right),\nonumber\\
\epsilon_\theta&\sim& \frac{Nu^{5/2}}{4K(RaPr)^{1/2}}\int_{1}^{\lambda_{BLT}/\lambda_u}dx\frac{x^2(1+2\eta-\eta^2 x^2)^2}{(1+x^2)^{3/2}(1+\eta^2x^2)^3},\quad\eta=\lambda_u/\lambda_\theta.
\label{fluctuest}
\ENA
Based on these expressions, we can singularize six different regimes. In the first three regimes, the dissipation is dominated by the mean flow contribution. This regimes are expected to hold at low Reynolds number (the turbulence is weak) or when the Prandtl number is low, according to the numerical study of \cite{VerzCamu99}. The thermal dissipation then fixes $\lambda_{BLT}\sim Nu^{-1}$. Then we can distinguish two cases: at very low Reynolds number, the boundaries are non-turbulent, and we have $\lambda_{BLV}<\lambda_u$, so that $\epsilon_U\sim u_\tau^4 \lambda_{BLV}/Pr$, and the two boundary layer scales follow (\ref{resume}) with $\epsilon=1$. In this regime, we get:
\EQ
Nu\sim Ra^{1/4} Pr^{-1/12},\quad Re\sim Ra^{3/8} Pr^{-5/8}.
\label{regime1}
\EN
This Nusselt versus Rayleigh dependence corresponds to regime $I_u$ of \cite{GrosLohs00}. They argue that for fixed Raleigh number, it could explain the weak decrease of the Nusselt number observed at large Prandtl number. For larger Reynolds number, when $\lambda_{BLT}>\lambda_u$, we get to a regime where $\epsilon_U\sim u_\tau^3\sim Ra Pr Nu$. This corresponds to a Reynolds number varying like $(Ra Nu)^{1/3}$, a situation observed in the low Prandtl number experiment of \cite{CCS97}. The final expression of $Nu$ depends on how the thermal boundary layer scale $\lambda_{BLT}\sim Nu^{-1}$ varies with the mean flow. Using (\ref{resume}), we obtain:
\EQ
Nu\sim \frac{Ra^{(1+\epsilon)/(5+2\epsilon)}Pr^{(1-2\epsilon)/(5+2\epsilon)}}{(1/\ln(Nu)+1/f)^{3\delta(\epsilon)/(5+2\epsilon)}}.
\label{regime2}
\EN
For $\epsilon\sim 1$ (which may be obtained at low Rayleigh, for $Pr>0.35$), this corresponds to the $2/7$ law of \cite{ShraSigg90}, in which the Nusselt decreases like $Pr^{-1/7}$ and the Reynolds number varies like $Ra^{-3/7}$. For low Prandtl, we have to consider the case $\epsilon=0$. In that case, we observe an interesting regime, where the Nusselt varies like $Ra^{1/5}(\ln Ra)^{3/5}$ at moderate Ra, then turning into a $Ra^{1/5}$ regime for larger Ra. The logarithmic correction increases the scaling exponent up to a value close to $1/4$, as shown in Fig. 1. In this regime, experimental measurements will tend to show first a 1/4 regime, followed by transition to a 1/5 regime, when the logarithmic corrections vanish, i.e. when the scaling of the boundary layer scale is determined by the bulk flow. This regime therefore exactly explain the experiments by \cite{CCS97}.\

Other interesting regimes appear when the fluctuations dominate the kinetic or thermal dissipation. The conditions under which this situation appears cannot be determined by a mere estimate of the ratio of $\epsilon_u$ versus $\epsilon_U$, for example, because there are some indications that this condition is in fact subject to boundary conditions. For example, experiments conducted at same aspect ratio, same cylindrical geometry and same Prandtl number by \cite{CCS97} and \cite{TSGS96} lead to different values of the Nusselt number. In \cite{CCS97}, the scaling relation $Re\sim (Ra Nu)^{1/3}$ (characterizing the regime in which kinetic energy dissipation is dominated by the mean flow ) is satisfied. In the experiment of \cite{TSGS96}, a careful study of the length scales revealed a significant deviation from the law $\lambda_{BLT}\sim 1/Nu$ which would be obtained in the regime explaining the experiment by \cite{CCS97}, and in which the thermal dissipation is provided by the mean temperature profile. It is then logical to attribute the difference of Nusselt via a difference in boundary conditions, which would favor or not the dominance of fluctuations. For example, the thickness or composition of the bottom plate may play such a role (Castaing, private communication).\

Let us consider first the regime in which mean flow dominates the thermal dissipation, but fluctuations dominates the kinetic dissipation. In such a case, we still have $\lambda_{BLT}\sim 1/Nu$, but the scaling of $Nu$ depends on the behavior of $\lambda_{BLV}$. Assuming (\ref{resume}), and taking $\epsilon=1$ for $\lambda_{BLT}$ (no fluctuations of temperature), we get:
\EQ
Nu\sim \frac{Ra^{(2+\epsilon)/(8+7\epsilon)}Pr^{\epsilon/(8+7\epsilon)}\left[\ln(\lambda_{BLV}/\lambda_u)\right]^{-2(2+\epsilon)/(8+7\epsilon)}}{(1/\ln(\lambda_{BLV})+1/f)^{-4\delta(\epsilon)/(8+7\epsilon)}}.
\label{regime3}
\EN
For $\epsilon=-1/2$ (valid for $Pr>0.35$), we obtain two different behaviour with (\ref{regime3}): one, with $\lambda_{BLV}/\lambda_u\approx 1$. This situation prevails for example under stress-free boundary conditions, like studied by \ {WernePC}, or for very large aspect ratio.  It may also occur at very large Prandtl number, if, as discussed in \cite{GrosLohs00} the viscous length scale approaches the size of the box and cannot increase any further. In such situation, we get $Nu=Ra^{1/3} Pr^{-1/9}$. In the opposite situation, where $\lambda_{BLV}/\lambda_u\gg 1$, we then have (within logarithmic corrections) $\lambda_{BLV}\sim Re^{1/3}$. This gives $Nu\sim Ra^{1/3}/(\ln(Ra))^{2/3}$. This law reproduces extremely well the recent experimental measurement of Niemela et al. \cite{NSSD00}, obtained with Helium ($Pr=0.7$) over 11 decades of Rayleigh numbers. These measurements can be fitted by a power-law with exponent $0.308$ over this range of Rayleigh number. However, as shown in Figure 2., even over such a wide range, the law with logarithmic corrections cannot be distinguished from the power-law exponent. Other measurements are also explained by this regime, for $Pr>0.35$. For example, in this regime, we predict a variation of the Nusselt number with the Prandtl numer going like $Nu\sim Pr^{-1/9}/\ln(Pr)$ (because $Re$ depends on the Prandtl number). This variation is shown in Figure. 3. A power-law fit of this variation over $0.7<Pr<100$ gives an exponent $-0.2$, exactly like in the experiment of Ashkenazi and Steinberg \cite{AshkStein99}. Also, the Reynolds number in this regime varies like $Re=\sqrt{Nu^3/Pr}$. The power-law dependence mimicked by the logarithmic corrections is (see Fig. 4 and 5) $Re\sim Ra^{0.46}Pr^{-0.8}$, again, in excellent agreement with the measurement of \cite{AshkStein99}. Other quantities, like the scaling of the central fluctuating temperature, or the length scales (Table IV) are also in excellent agreement with direct measurements of Niemela et al. \cite{NSSD00} or by Kerr \cite{Kerr96}. We do not discuss the regime corresponding to $\epsilon=0$ in (\ref{regime3}) because, apparently, it is not applicable to the low Prandtl regime: it assumes that $\lambda_{BLT}<\lambda_u$, at variances with experimental observations by Segawa et al \cite{SNS98}.\

The last regime is obtained when temperature fluctuations dominate the thermal dissipation, making $\lambda_{BLT}$ differ from $1/Nu$. The two interesting regimes then depend on whether $\eta=\lambda_u/\lambda_\theta=Nu/Re$ is less or larger than $1$. In the first case, we get:
\EQA
\epsilon_\theta&\sim&\frac{Nu^{5/2}}{4K(RaPr)^{1/2}}\int_{1}^{\lambda_{BLT}/\lambda_u}dx\frac{x^2}{(1+x^2)^{3/2}}\nonumber\\
&\sim& \frac{Nu^{5/2}}{4K(RaPr)^{1/2}}\ln\left(\lambda_{BLT}/\lambda_u)\right).
\label{approxetapetit}
\ENA
We then get from (\ref{exacts}) that $Nu\sim (Ra Pr)^{1/3}/(\ln(\lambda_{BLT}/\lambda_u))^{2/3}$. This regime is similar to regime (\ref{regime3}), with a different prefactor. In the case where $\eta\gg 1$, we get
\EQA
\epsilon_\theta&\sim&\frac{Nu^{5/2}}{4K(RaPr)^{1/2}}\int_{1}^{\lambda_{BLT}/\lambda_u}\frac{dx}{\eta^2 x^2}\nonumber\\
&\sim& \frac{Nu^{1/2}u_\tau^2}{4K(RaPr)^{1/2}}.
\label{approxetagrand}
\ENA
From (\ref{exacts}), we  get $Re=u_\tau/Pr=Pr^{-3/4} (Ra Nu)^{1/4}$. Using the expression (\ref{resume}) for $\lambda_{BLV}$, and $\epsilon_u=Ra Pr Nu$, we finally get :
\EQ
Nu\sim \frac{Ra^{1/(3+2\epsilon)}Pr^{(1+2\epsilon)/(3+2\epsilon)} \ln(\lambda_{BLV}/\lambda_u)^{-2(2+\epsilon)/(3+2\epsilon)}}{(1/\ln(\lambda_{BLV})+1/f)^{-4\delta(\epsilon)/(3+2\epsilon)}}.
\label{regime5}
\EN
For $\epsilon=0$ (corresponding to low Prandtl number), we obtain a regime $Nu=Ra^{1/3}$ with logarithmic corrections which depend on the Rayleigh number: for low Rayleigh number, the scaling of the boundary layer is determined by the logarithmic region, making $Nu\sim Ra^{1/3}/\ln(Ra)^{8/3}$. This regime mimics a scaling exponent close to $1/4$ (see Figure 6). For larger Rayleigh numbers, the scaling of the boundary layer scale is determined by the bulk flow, changing the correction to $Nu\sim Ra^{1/3}/(\ln(Ra))^{4/3}$. This results in a larger effective exponent, of the order of $0.28$. In this regime, the boundary layer scale like $\lambda_{BLV}\sim (Re\ln(Re))^{-1/2}$, providing an approximate power-law in Rayleigh with exponent $-0.2$. All these features were observed in the mercury experiment of \cite{GSNS99}. For $\epsilon=-1/2$ ($Pr>0.35$, we obtain a regime in which the Nusselt depends only weakly on the Prandtl number
$Nu\sim Ra^{1/2}/(\ln(Re))^{3/2}$. This regime corresponds to the "ultra-hard" convective regime predicted by Kraichnan \cite{Krai62}, but with logarithmic corrections. In Fig. 7, we compare this regime with the $Nu\sim Pr^{0.072}Ra^{0.389}$ approximate power-law measured by Chavanne et al \cite{CCCHCC97} in Helium, at $Ra>10^{11}$. Our formula predicts a very weak dependence of in the Prandtl number, like in  \cite{CCCHCC97}, but with opposite sign. This could be accounted for minute variations of the mean profile around the value $\epsilon=-1/2$. However, the plus sign of the fit by Chavanne et al. could be an artifact due to the scatter of the data.  Indeed several measurements performed at fixed Rayleigh number and for increasing Prandtl show that $Nu$ tends to decrease rather than increase, as in the formula by Chavanne et al. The predictions in this regime are also in agreement with their Reynolds number measurements $Re\sim Ra^{0.49} Pr^{-0.75}$ if one considers that this Reynolds number is based on the vertical velocity, i.e. $Re\sim w_c Pr$. Interestingly enough, the slight discrepancy between our prediction for $d\ln w_c/d\ln Pr$ (little less than 0.25) and the Chavanne et al. fit (little more than 0.25) can be traced back to the discrepancy in the Nusselt number. So there must be some systematic error present at this level, either in the theory, or in the experiments. In any case, this new regime may explain to puzzling facts: 1) why Glazier et al do not observe transition towards a new scaling regime, despite their very large Reynolds number. This is because they were already in the ultra-hard regime! 2) why for the same geometry and the same aspect ratio, Niemela et al did not observe a transition towards the regime observed by Chavanne et al. It is because, somehow, they do not allow the temperature fluctuations to grow enough so that their contribution supersedes that of the mean flow.\

A good summary of our results can be found in the tables. For completeness, we have indicated the scaling behavior of various quantities measured in experiments, like $\Delta_c$ or $w_c$ (the value of temperature or velocity fluctuations at the center of the cell). These values can be estimated from (\ref{oufo}), with $z=1$. When available, we have indicated comparison with  measurements. This summary shows that many scaling behaviors observed in recent experiments can be reproduced using log-corrected scaling laws. Overall, the agreement of the theory based our toy model of thermal convection is excellent for the case $Pr\sim 1$ (Helium). It predicts reasonably well the Rayleigh dependence of the quantities for low Prandtl number, but it is not clear whether the Prandtl number dependence in these regimes is well predicted. This might be caused by our initial hypothesis (taking $Pr^t=1$). Also, we were unsure where to put in this picture the recent experiments by \cite{XBA00} performed in water. The difficulty is that these experiments are performed at many different aspect ratio, and it is difficult to disentangle  the aspect ratio dependence of the local scaling exponents with other dependences. At least in the case of Helium, apparently, we reached a point where experimental measurements cannot discriminate between logarithmic corrections to scaling, or corrections due to superposition of power laws, like  proposed by \cite{GrosLohs00}. This calls for identification of more stringent tests of theories. One possibility, which has been so far under-exploited, would be to concentrate on statistical properties rather than average, i.e. on probability distributions functions. In this context, it is interesting to note that our toy model of turbulent convection allows such investigations, via a Langevin formalism exploiting the linearity and the stochasticity of the equations. In neutral condition, our toy model leads to PDF's for velocity with qualitatively similar behavior than the PDF's measured in high Reynolds number flows \cite{LDN00}. The generalization of these results to the convective regime is underway.

I thank the Fluid Mechanics group of the ENS Lyon for many very interesting discussions. My special thanks goes to B. Castaing for his continuous interest and support in the present work.

\begin{table}
\begin{tabular}{c|c|c|c|c}\hline
\multicolumn{1}{c|}{Q(X,Y)} &\multicolumn{2}{c|}{This theory}
&\multicolumn{2}{c}{Experiments}\\
\multicolumn{1}{c|}{}&\multicolumn{1}{c|}{$d\ln Q/dX$}
&\multicolumn{1}{c|}{$d\ln Q/dY$} &\multicolumn{1}{c|}{$d\ln Q/dX$}
&\multicolumn{1}{c}{$d\ln Q/dY$}\\
\hline
$Nu$ &$1/4$ &$-1/12$ &$$
&$$ \\
$Re=u_\tau/Pr$ &$3/8$ &$-5/8$ &$$
&$$ \\
$\Delta_c$ &$-3/16$ &$-13/48$ &$$ &$$\\
$w_c$ &$7/16$ &$3/16$ &$$ &$$\\
$\lambda_{BLT}$ &$-1/4$ &$1/12$ &$$
&$$\\
$\lambda_{BLV}$ &$-1/4$ &$5/12$ &$$
&$$ \\\hline
\end{tabular}
\label{table1}
\caption{Summary of local exponents of different physical quantities in regime 1: low Reynolds number, mean flow dominates. In this table $X=\ln Ra$ and $Y=\ln Pr$. No dependence on $X$ or $Y$ of the scaling exponent indicates real scaling with respect to $Ra$ or $Pr$ respectively.}
\end{table}

\begin{table}
\begin{tabular}{c|c|c|c|c}\hline
\multicolumn{1}{c|}{Q(X,Y)} &\multicolumn{2}{c|}{This theory}
&\multicolumn{2}{c}{Experiments}\\
\multicolumn{1}{c|}{}&\multicolumn{1}{c|}{$d\ln Q/dX$}
&\multicolumn{1}{c|}{$d\ln Q/dY$} &\multicolumn{1}{c|}{$d\ln Q/dX$}
&\multicolumn{1}{c}{$d\ln Q/dY$}\\
\hline
$Nu$ &$2/7$ &$-1/7$ &$0.282$
&$$ \\
$Re=u_\tau/Pr$ &$3/7$ &$-5/7$ &$0.43$
&$$ \\
$\Delta_c$ &$-5/28$ &$-2/7$ &$-0.147$ &$$\\
$w_c$ &$13/28$ &$1/7$ &$0.491$ &$$\\
$\lambda_{BLT}$ &$2/7$ &$-1/7$ &$$
&$$\\
$\lambda_{BLV}$ &$np$ &$np$ &$$
&$$ \\\hline
\end{tabular}
\label{table2}
\caption{Summary of local exponents of different physical quantities in regime 2: $Pr>0.35$, mean flow dominates. This regime corresponds to the theory of 
\protect\cite{ShraSigg90}. The measurements are those of the Chicago group 
\protect\cite{Cast89}. However, this experiments might well be better described by regime 4, see below. 
In this table $X=\ln Ra$ and $Y=\ln Pr$. No dependence on $X$ or $Y$ of 
the scaling exponent indicates real scaling with respect to 
$Ra$ or $Pr$ respectively. The symbol $np$ means not predicted by this theory.
}
\end{table}

\begin{table}
\begin{tabular}{c|c|c|c|c}\hline
\multicolumn{1}{c|}{Q(X,Y)} &\multicolumn{2}{c|}{This theory}
&\multicolumn{2}{c}{Experiments}\\
\multicolumn{1}{c|}{}&\multicolumn{1}{c|}{$d\ln Q/dX$}
&\multicolumn{1}{c|}{$d\ln Q/dY$} &\multicolumn{1}{c|}{$d\ln Q/dX$}
&\multicolumn{1}{c}{$d\ln Q/dY$}\\
\hline
$Nu$ &$1/5+3f/5(fX+X^2)$ &$1/5+3f/5(fY+Y^2)$ &$0.26$ then $0.2$
&$0.21$ \\
$Re=u_\tau/Pr$ &$2/5+f/5(fX+X^2)$ &$-3/5+f/5(fY+Y^2)$ &$0.424$
&$$ \\
$\Delta_c$ &$-1/5+3f/20(fX+X^2)$ &$-1/5+3f/20(fY+Y^2)$ &$$ &$$\\
$w_c$ &$2/5+9f/20(fX+X^2)$ &$2/5+9f/20(fY+Y^2)$ &$$ &$$\\
$\lambda_{BLT}$ &$-1/5-3f/5(fX+X^2)$ &$-1/5-3f/5(fY+Y^2)$ &$$
&$$\\
$\lambda_{BLV}$ &$np$ &$np$ &$$
&$$ \\\hline
\end{tabular}
\label{table3}
\caption{Summary of local exponents of different physical quantities in regime 3: $Pr<0.35$, mean flow dominates. The measurements  are from 
\protect\cite{CCS97}. In this table $X=\ln Ra$ and $Y=\ln Pr$ and $f$ is a constant, depending on the shape of the bulk velocity profile. This constant is not predicted by the theory. No dependence on $X$ or $Y$ of the scaling exponent indicates real scaling with respect to $Ra$ or $Pr$ respectively. The symbol $np$ means not predicted by this theory.}
\end{table}

\begin{table}
\begin{tabular}{c|c|c|c|c}\hline
\multicolumn{1}{c|}{Q(X,Y)} &\multicolumn{2}{c|}{This theory}
&\multicolumn{2}{c}{Experiments}\\
\multicolumn{1}{c|}{}&\multicolumn{1}{c|}{$d\ln Q/dX$}
&\multicolumn{1}{c|}{$d\ln Q/dY$} &\multicolumn{1}{c|}{$d\ln Q/dX$}
&\multicolumn{1}{c}{$d\ln Q/dY$}\\
\hline
$Nu$ &$1/3-2/(3X)$ &$-1/9-2/3Y$ &$0.309^1$
&$-0.2^2$ \\
$Re=u_\tau/Pr$ &$1/2-1/X$ &$-2/3-1/Y$ &$0.43^{2,3}$
&$-0.75^2$ \\
$\Delta_c$ &$-1/6-1/6X$ &$-5/18-1/6Y$ &$-0.145^1$ &$$\\
$w_c$ &$1/2-1/2X$ &$1/6-1/2Y$ &$0.53^4$ &$0.08^4$\\
$\lambda_{BLT}$ &$-1/3+2/3X$ &$1/9+2/3Y$ &$-0.28^3$
&$$\\
$\lambda_{BLV}$ &$-1/6+1/3X$ &$1/9+1/3Y$ &$-0.14^3$
&$0.21^3$ \\\hline
\end{tabular}
\label{table4}
\caption{Summary of local exponents of different physical quantities in regime 4: $Pr>0.35$, fluctuations dominates. The measurements  are from 
\protect\cite{CCS97} (superscript 1), 
\protect\cite{AshkStein99} (superscript 2), 
\protect\cite{KerrHerr00} (superscript 3) and \protect\cite{VerzCamu99}
 (superscript 4). In this table $X=\ln Ra$ and $Y=\ln Pr$. No dependence on $X$ or $Y$ of the scaling exponent indicates real scaling with respect to $Ra$ or $Pr$ respectively. The symbol $np$ means not predicted by this theory.}
\end{table}

\begin{table}
\begin{tabular}{c|c|c|c|c}\hline
\multicolumn{1}{c|}{Q(X,Y)} &\multicolumn{2}{c|}{This theory}
&\multicolumn{2}{c}{Experiments}\\
\multicolumn{1}{c|}{}&\multicolumn{1}{c|}{$d\ln Q/dX$}
&\multicolumn{1}{c|}{$d\ln Q/dY$} &\multicolumn{1}{c|}{$d\ln Q/dX$}
&\multicolumn{1}{c}{$d\ln Q/dY$}\\
\hline
$Nu$ &$1/3-4/3X-4f/3(fX+X^2)$ &$1/3-4/(3Y)-4f/3(fY+Y^2)$ &$0.25$ then $0.285$
&$$ \\
$Re=u_\tau/Pr$ &$1/3-1/3X-f/3(fX+X^2)$ &$-2/3-1/3Y-f/3(fY+Y^2)$ &$$
&$$ \\
$\Delta_c$ &$-1/6-1/3X-f/3(fX+X^2)$ &$-1/6-1/3Y-f/3(fY+Y^2)$&$$ &$$\\
$w_c$ &$1/2-1/3X-f/3(fX+X^2)$ &$1/2-1/3Y-f/3(fY+Y^2)$ &$$ &$$\\
$\lambda_{BLT}$ &$np$ &$np$ &$$
&$$\\
$\lambda_{BLV}$ &$-1/6+1/6X+f/6(fX+X^2)$ &$-1/6+1/6Y+f/6(fY+Y^2)$ &$-0.2$
&$$ \\\hline
\end{tabular}
\label{table5}
\caption{Summary of local exponents of different physical quantities in regime 5: $Pr<0.35$, fluctuations dominates and $\lambda_\theta\gg\lambda_u$. The measurements  from \protect\cite{GSNS99}. In this table $X=\ln Ra$ and $Y=\ln Pr$ and $f$ is a constant, depending on the shape of the bulk velocity profile. This constant is not predicted by the theory. No dependence on $X$ or $Y$ of the scaling exponent indicates real scaling with respect to $Ra$ or $Pr$ respectively. The symbol $np$ means not predicted by this theory.}
\end{table}

\begin{table}
\begin{tabular}{c|c|c|c|c}\hline
\multicolumn{1}{c|}{Q(X,Y)} &\multicolumn{2}{c|}{This theory}
&\multicolumn{2}{c}{Experiments}\\
\multicolumn{1}{c|}{}&\multicolumn{1}{c|}{$d\ln Q/dX$}
&\multicolumn{1}{c|}{$d\ln Q/dY$} &\multicolumn{1}{c|}{$d\ln Q/dX$}
&\multicolumn{1}{c}{$d\ln Q/dY$}\\
\hline
$Nu$ &$1/2-3/(2X)$ &$-3/2Y$ &$0.389$
&$0.072$ \\
$Re=u_\tau/Pr$ &$3/8-3/8X$ &$-3/4-3/8Y$ &$$
&$$ \\
$\Delta_c$ &$-1/8-3/(8X)$ &$-1/4-3/8Y$ &$$ &$$\\
$w_c$ &$5/8-9/8X$ &$1/4-9/8Y$ &$0.49$ &$0.28$\\
$\lambda_{BLT}$ &$np$ &$np$ &$$
&$$\\
$\lambda_{BLV}$ &$-1/8+1/8X$ &$1/4+1/4Y$ &$$
&$$ \\\hline
\end{tabular}
\label{table6}
\caption{Summary of local exponents of different physical quantities in regime 6: $Pr>0.35$, fluctuations dominates and $\lambda_\theta\gg\lambda_u$. The measurements  are from \protect\cite{CCCHCC97}. 
In this table $X=\ln Ra$ and $Y=\ln Pr$. No dependence on $X$ or $Y$ of the scaling exponent indicates real scaling with respect to $Ra$ or $Pr$ respectively. The symbol $np$ means not predicted by this theory.}
\end{table}

\newpage
\epsffile{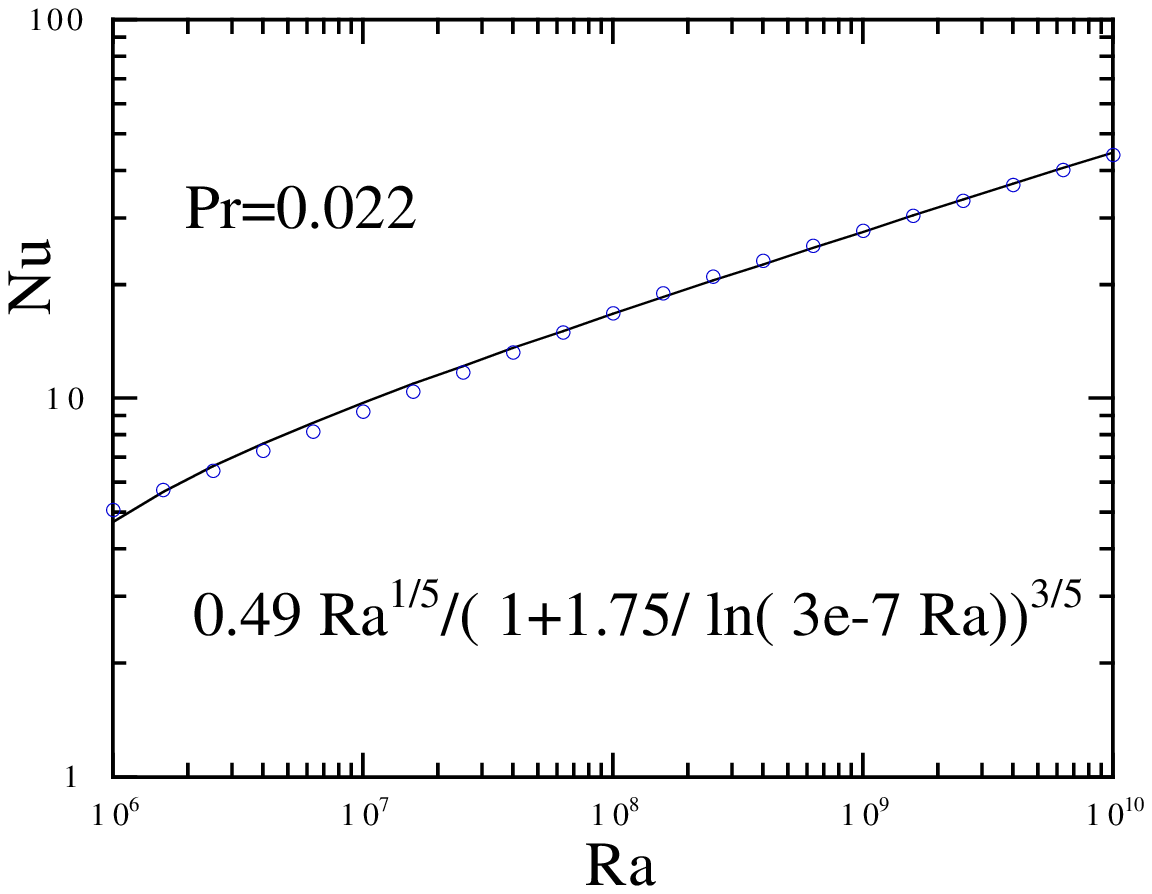}
\begin{figure}[hhh]
\caption[]{ Nusselt vs Rayleigh in regime 3 ($Pr<0.35$, mean flow dominates). The symbols are the power-law fits of the experimental measurements by 
\protect\cite{CCS97}. The line is the theoretical formula predicted by the toy model (eq. (\ref{regime2}) with $\epsilon=0$).}
\label{fig:fig1}
\end{figure}

\newpage
\epsffile{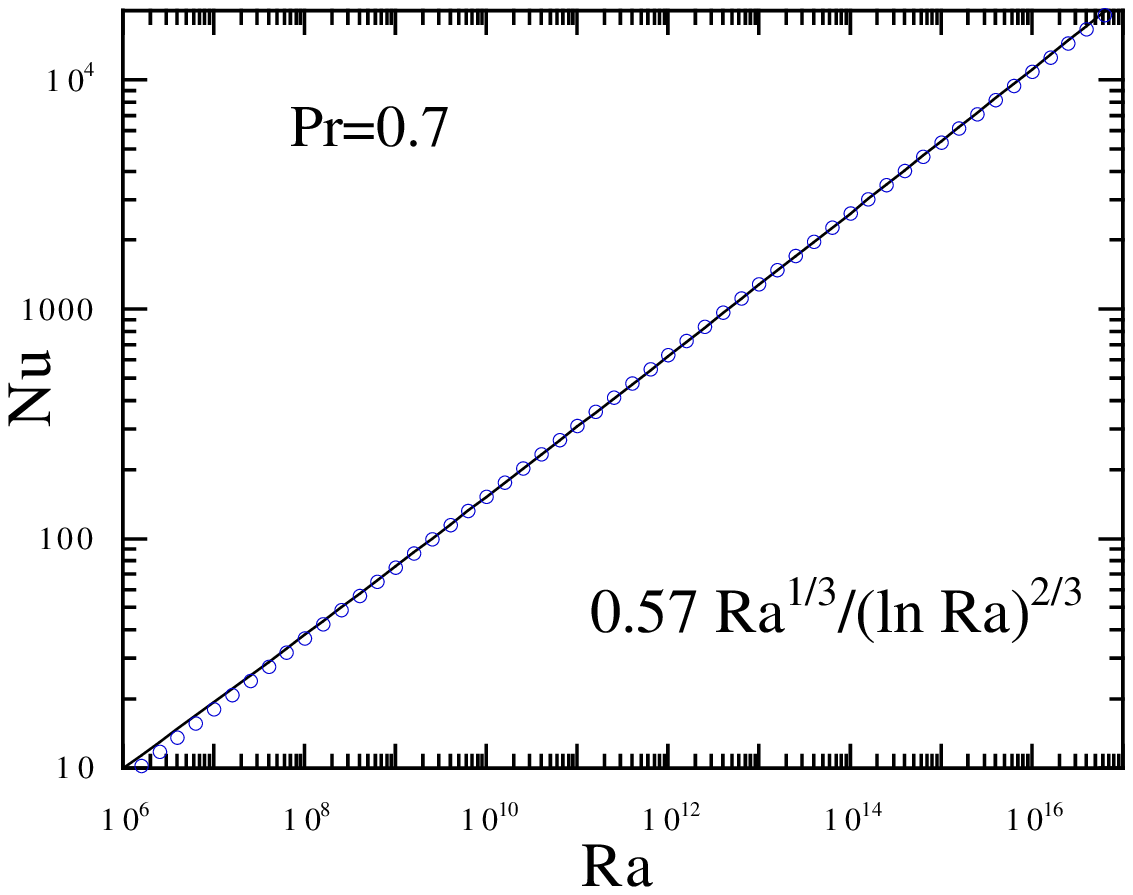}
\begin{figure}[hhh]
\caption[]{Nusselt vs Rayleigh in regime 4 ($Pr>0.35$, velocity fluctuations dominate but temperature fluctuations are negligible). The symbols are the power-law fit $Nu\sim Ra^{0.309}$  of the experimental measurements by 
\protect\cite{NSSD00}. The line is the theoretical formula predicted by the toy model (eq. (\ref{regime3}) with $\epsilon=-1/2$).}
\label{fig:fig2}
\end{figure}

\newpage
\epsffile{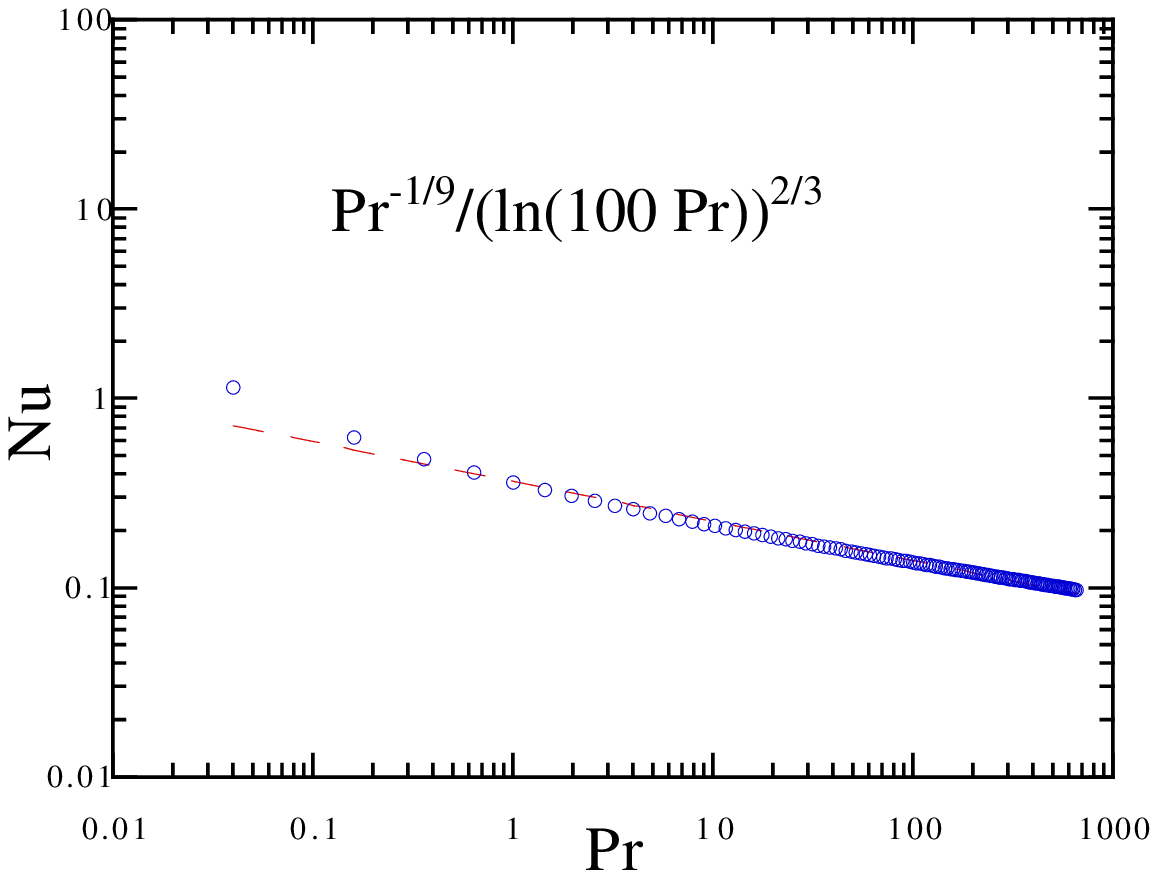}
\begin{figure}[hhh]
\caption[]{Nusselt vs Prandtl in regime 4 ($Pr>0.35$, velocity fluctuations dominate but temperature fluctuations are negligible). The symbols is the theoretical formula (eq. (\ref{regime3}) with $\epsilon=-1/2$). The dotted line is a power-law fit, with slope $-0.2$, mimicking the experimental fit of 
\protect\cite{AshkStein99}.}
\label{fig:fig3}
\end{figure}

\newpage
\epsffile{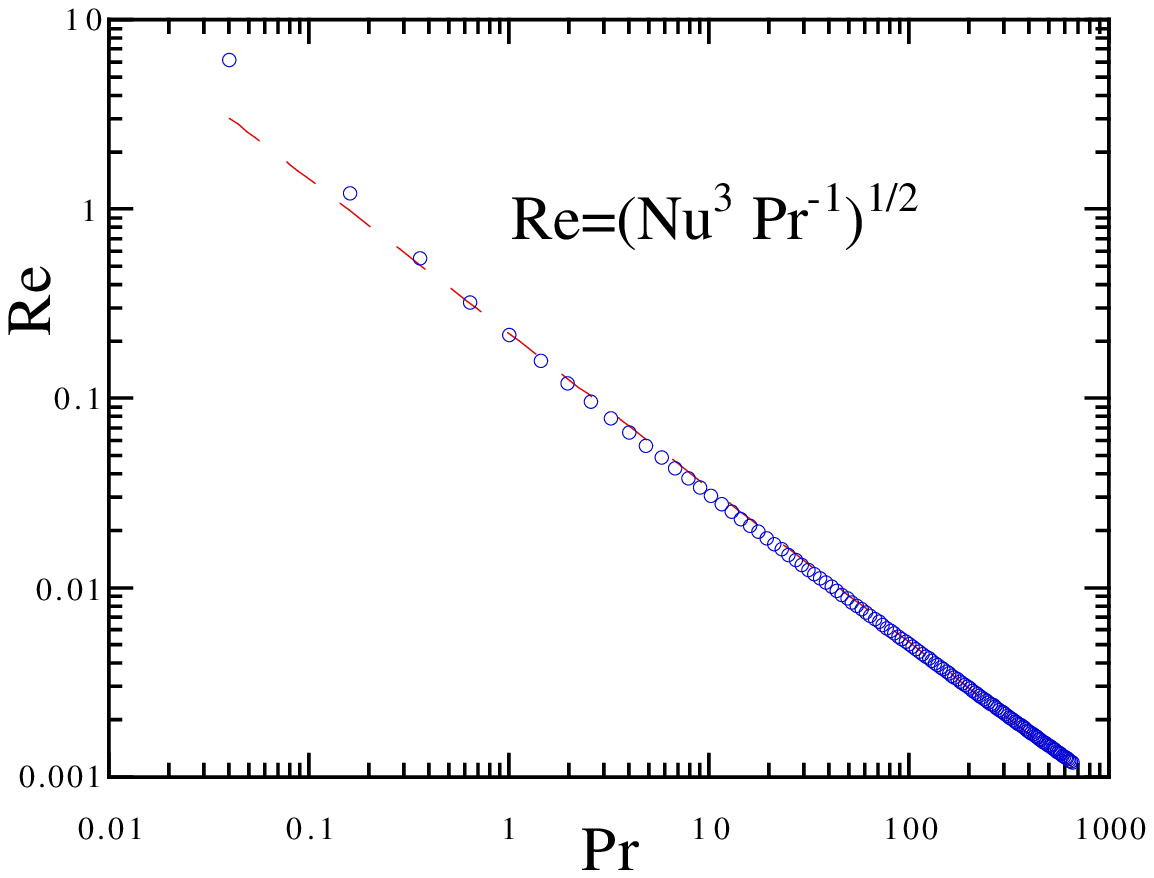}
\begin{figure}[hhh]
\caption[]{ Reynolds  vs Prandtl in regime 4 ($Pr>0.35$, velocity fluctuations dominate but temperature fluctuations are negligible). The symbols is the theoretical formula. The dotted line is a power-law fit, with slope $-0.8$, mimicking the experimental fit of \protect\cite{AshkStein99}.}
\label{fig:fig4}
\end{figure}

\newpage
\epsffile{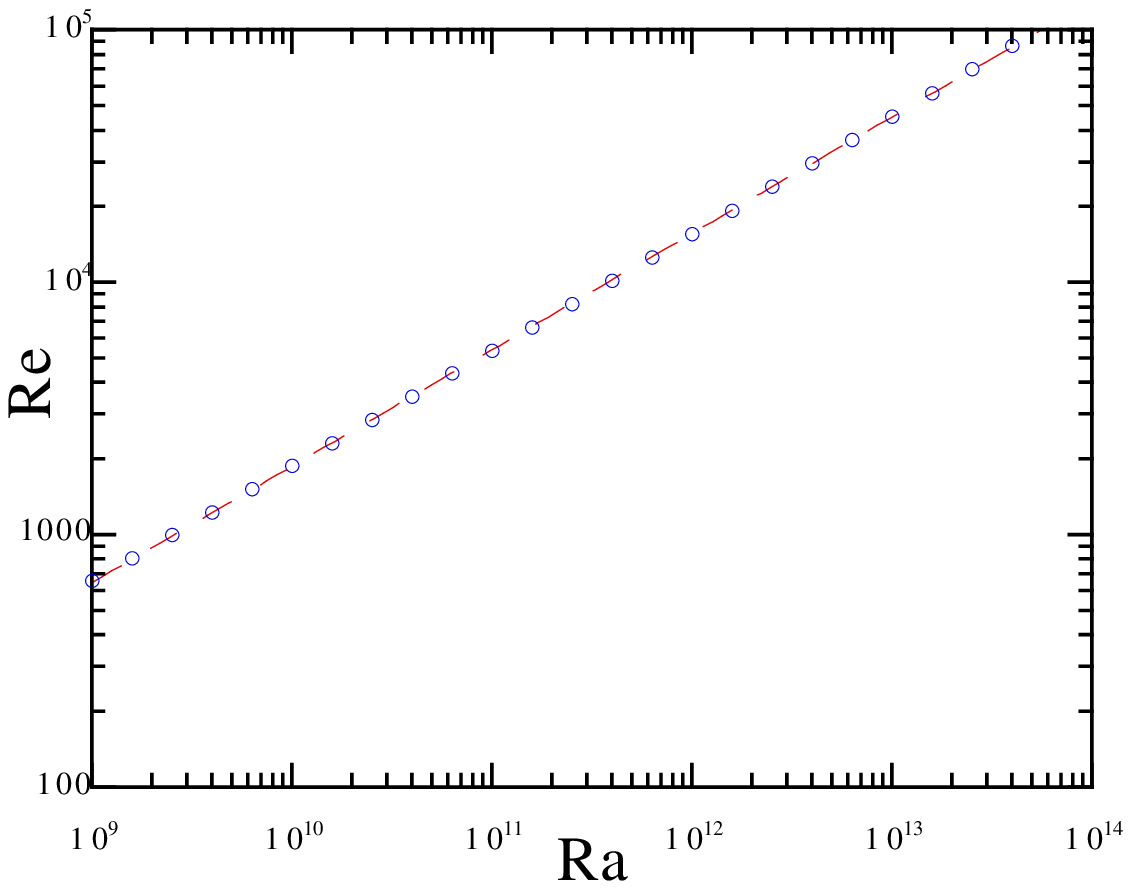}
\begin{figure}[hhh]
\caption[]{Reynolds vs Rayleigh in regime 4 ($Pr>0.35$, velocity fluctuations dominate but temperature fluctuations are negligible). The symbols is the theoretical formula. The dotted line is a power-law fit, with slope $0.43$, mimicking the experimental fit of \protect\cite{AshkStein99}.}
\label{fig:fig5}
\end{figure}

\newpage
\epsffile{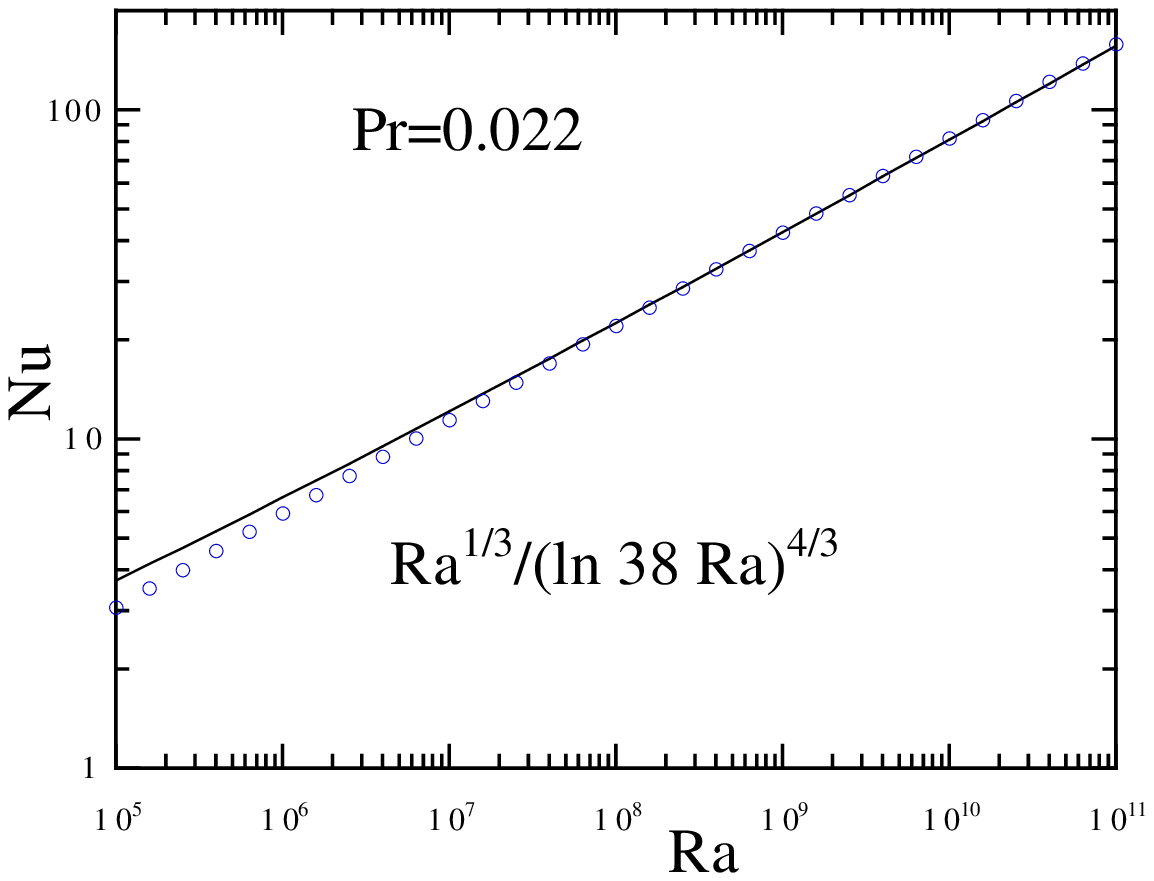}
\begin{figure}[hhh]
\caption{ Nusselt vs Rayleigh in regime 5 ($Pr<0.35$, fluctuations dominates. The symbols are the power-law fits of experimental measurements by 
\protect\cite{GSNS99}. The line is the theoretical formula predicted by the toy model (eq. (\ref{regime5}) with $\epsilon=0$).}
\label{fig:fig6}
\end{figure}

\newpage
\epsffile{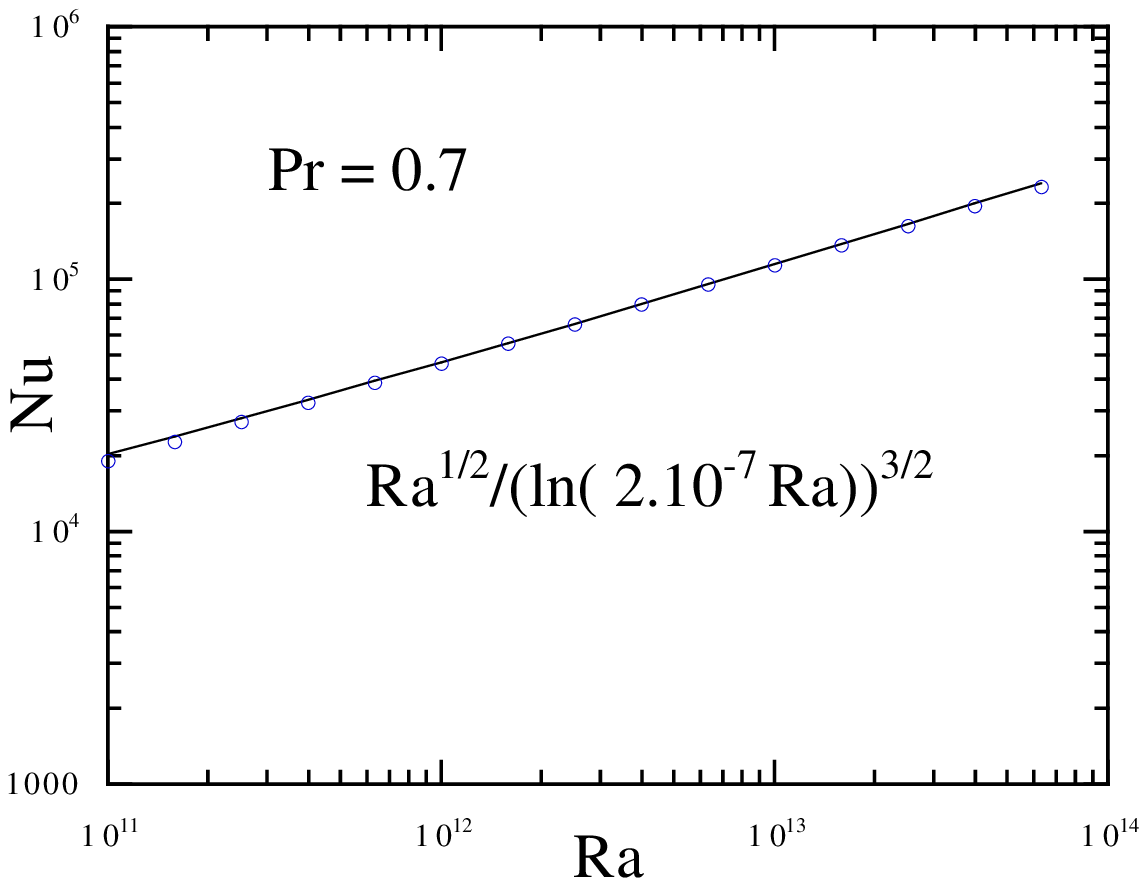}
\begin{figure}[hhh]
\caption[]{ Nusselt vs Rayleigh in regime 6 ($Pr>0.35$, fluctuations dominates. The symbols are the power-law fits of the experimental measurements by 
\protect\cite{CCCHCC97}. The line is the theoretical formula predicted by the toy model (eq. (\ref{regime5}) with $\epsilon=-1/2$).}
\label{fig:fig7}
\end{figure}


\begin{thebibliography}{99}

\bibitem{Gols}
R.J. Goldstein, H.D. Chiang and D.L. See, {\sl J. Fluid Mech.}
{\bf 213}, 111 (1990)

\bibitem{CCS97}
S. Cioni, S. Ciliberto and J. Sommeria {\sl J. Fluid Mech.}
{\bf 335} 111 (1997)

\bibitem{Cast89}
B. Castaing, G. Gunaratne, F. Heslot, L. Kadanoff, A. Libchaber,S.
Thomae,X-Z. Wu, S. Zaleski and G. Zanetti
{\sl J. Fluid Mech.} {\bf 204} 1 (1989).

\bibitem{ShraSigg90}
B. I. Shraiman and E.D. Siggia 
{\sl Phys. Rev. A.} {\bf 42}, 3650 (1990).

\bibitem{Krai62}
R. Kraichnan {\sl Phys. Fluids} {\bf 5} 1374 (1962)

\bibitem{GrosLohs00}
S. Grossmann and D. Lohse {\sl J. Fluid Mech.} {\bf 407} 27 (2000).

\bibitem{GrosLohs01}
S. Grossman and D. Lohse preprint (2000).

\bibitem{XBA00}
X. Xu, K.M.S. Bajaj and G. Ahlers, {\sl Phys. Rev. Letter}, {\bf 84}, 4357 (2000).

\bibitem{Malk54}
W.V.R. Malkus 
{\sl Proc. Roy. Soc. London., A} {\bf 225}, 185-195. (1954);
W.V.R. Malkus 
{\sl Proc. Roy. Soc. London., A} {\bf 225}, 195-212. (1954)

\bibitem{Buss78}
F.H. Busse {\sl Rep. Prog. Phys. } {\bf 41} 1930 (1978).

\bibitem{NSSD00}
J.J. Niemela, L. Skrbek, K.R. Sreenivasan and R.J. Donnelly
 {\sl Nature} {\bf 404} 837 (2000).

\bibitem{DLS00}
B. Dubrulle, J-P. Laval and P. Sullivan preprint available at 
http://webast.ast.obs-mip.fr/people/bdubru. 
(2000).

\bibitem{Dubr00}
B. Dubrulle, {\sl Europhys. Letters} {\bf 51} 513 (2000).



\bibitem{DLSW00}
B. Dubrulle, J-P. Laval, P. Sullivan and J. Werne A dynamical subgrid scale model for the planetary surface layer 1. The model {\sl submitted to J. Atmosph. Sci.} (2000).

\bibitem{Wern93}
J. Werne, 
{\sl Phys. Rev. E} {\bf 48} 1020-1035 (1993).



\bibitem{Naza99}
S. Nazarenko {\sl Phys. Lett. A} {\bf 264} 444 (2000);
S. Nazarenko, N. Kevlahan and B. Dubrulle 
{\sl Physica D} {\bf 139} 158 (2000);
B. Dubrulle, J-P. Laval, S. Nazarenko and N. Kevlahan  
preprint. 

\bibitem{SNS98}
T. Segawa, A. Naert and M. Sano, {\sl Phys. Rev. E} {\bf 57} 557 (2998); A. Naert, T. Segawa and M. Sano {\sl Phys. Rev. E} {\bf 56} R1302 (1997).


\bibitem{VerzCamu99}
R. Verzicco and R. Camussi {\sl J. Fluid Mech.} {\bf 383} 55 (1999)

\bibitem{Kerr96}
R.M. Kerr, {\sl J. Fluid Mech.} {\bf 310} 139 (1996). 

\bibitem{KerrHerr00}
R. M. Kerr and J. R. Herring (2000) preprint.

\bibitem{KerrPC}
R.M. Kerr, private communication (1999).


\bibitem{WernPC}
J. Werne, private communication; K. Julien, S. Legg, J. Mc Williams and 
J. Werne, {\sl J. Fluid Mech.} {\bf 322} 243 (1996).

\bibitem{CiliLaro99}
S. Ciliberto and C. Laroche {\sl Phys. Rev. Letters} {\bf 82} 3998 (1999).

\bibitem{CCL96}
S. Ciliberto, S. Cioni and C. Laroche {\sl Phys. Rev E} {\bf 54} R5901 (1996).

\bibitem{TSGS96}
T. Takeshita, T. Segawa, J.A. Glazier and M. Sano {\sl Phys. Rev. Letter}
{\bf 76} 1465 (1996).

 
\bibitem{AshkStein99}
S. Ashkenazi and V. Steinberg {\sl Phys. Rev. Letter} {\bf 83},
3641 (1999).

\bibitem{GSNS99}
J.A. Glazier, T. Segawa, A. Naert and M. Sano, {\sl Nature}
{\bf 398} 307 (1999).

\bibitem{CCCHCC97}
X. Chavanne, F. Chilla, B. Castaing, B. H\'ebral, B. Chabaud and
J. Chaussy, {\sl Phys. Rev. Letters} {\bf 79}, 3648 (1997).


\bibitem{LDN00}
J-P. Laval, B. Dubrulle and S. Nazarenko Non-locality and intermittency in 3D turbulence {\sl submitted to Phys. Fluids} (2000).

\end{thebibliography}
\end{document}